\pdfoutput=1
\documentclass[conference]{IEEEtran}
\usepackage{cite}
\usepackage{amsmath,amssymb,amsfonts}
\usepackage{dsfont}
\usepackage{algpseudocode}
\usepackage{graphicx}
\usepackage{textcomp}
\def\BibTeX{{\rm B\kern-.05em{\sc i\kern-.025em b}\kern-.08em
    T\kern-.1667em\lower.7ex\hbox{E}\kern-.125emX}}
\usepackage{shortcuts}

\graphicspath{{./}}

\title{Parallel Contextual Bandits in Wireless Handover Optimization}

\author{\IEEEauthorblockN{Igor Colin}
\IEEEauthorblockA{\textit{Noah's Ark Lab Paris} \\
\textit{Huawei Technologies Ltd}\\
igor.colin@huawei.com}
\and
\IEEEauthorblockN{Albert Thomas}
\IEEEauthorblockA{\textit{Noah's Ark Lab Paris} \\
\textit{Huawei Technologies Ltd}\\
albert.thomas@huawei.com}
\and
\IEEEauthorblockN{Moez Draief}
\IEEEauthorblockA{\textit{Noah's Ark Lab Paris} \\
\textit{Huawei Technologies Ltd}\\
moez.draief@huawei.com}
}

\begin{document}
\maketitle

\begin{abstract}
  As cellular networks become denser, a scalable and dynamic tuning of wireless base station parameters can only be achieved through automated optimization. Although the contextual bandit framework arises as a natural candidate for such a task, its extension to a parallel setting is not straightforward: one needs to carefully adapt existing methods to fully leverage the multi-agent structure of this problem. We propose two approaches: one derived from a deterministic UCB-like method and the other relying on Thompson sampling. Thanks to its bayesian nature, the latter is intuited to better preserve the exploration-exploitation balance in the bandit batch. This is verified on toy experiments, where Thompson sampling shows robustness to the variability of the contexts. Finally, we apply both methods on a real base station network dataset and evidence that Thompson sampling outperforms both manual tuning and contextual UCB.
\end{abstract}


\section{Introduction}
\label{sec:introduction}

\begin{figure}[ht!]
\centering
\includegraphics[width=7cm]{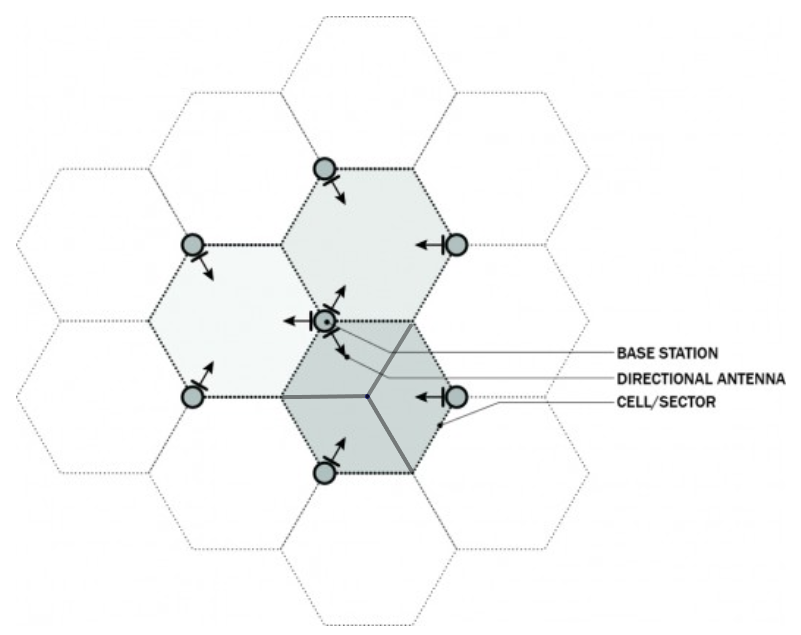}
\caption{Cellular network organization}
\label{fig:cellular_network_with_cells}
\end{figure}

The land area covered by a cellular wireless network, such as a mobile phone network, is divided into small areas called \emph{cells}, each cell being covered by the antenna of a fixed \emph{base station} (see Figure \ref{fig:cellular_network_with_cells}). Each base station is configured by a set of parameters that should be tuned so as to provide the best possible network coverage. Although default recommended values can be used, the best values of these parameters are likely to depend on the traffic (\eg~the number of users) and the geographical location of the base stations. Furthermore, these parameters often need to be adjusted on a regular basis in order to adapt to the evolution of the traffic. The manual tuning of base station parameters may thus be highly time consuming, tedious and needs to preserve some level of quality of service. In addition, recent developments in cellular network standards lean towards a densification of base stations, encouraging operators to find automated solutions for optimal parameters configuration (see \eg \cite{Siomina2006,awada2011taguchioptimization,capdevielle2013handover,isa2015handover}).

One possible way of modeling this problem is through contextual bandits \cite{lu2010contextual}: in this framework, one aims at optimizing an objective that is depending on contextual features (\eg traffic and environment) while avoiding too much deterioration of the objective function (\eg quality of service).
More specifically, at each time $t$, given side information about the current state of the wireless network (the \emph{context}), the operator wants to choose the values of the parameters (an \emph{action}) so as to obtain the best user experience (the \emph{reward}). However, since the relation between an action and its associated reward is initially unknown, one needs to \emph{explore} the space of available actions in order to gain some knowledge about this relation before being able to \emph{exploit} it. This exploration-exploitation tradeoff is common to every bandit problem, from multi-armed bandits \cite{lai1985asymptotically, auer2002finite, bubeck2009online} to lazy optimization through gaussian processes \cite{srinivas2009gaussian, contal2013parallel, desautels2014parallelizing, pedregosa2016hyperparameter}.

Contextual bandit problems have received a lot a attention in the past decade, either for theoretical guarantees \cite{chu2011contextual, agrawal2013thompson}, delayed rewards framework \cite{joulani2013online} or pure exploration scenarios \cite{audibert2010best, xu2017fully}. Surprisingly, the multi-agent setting, such as the one induced by the base station parameter tuning, has hardly been investigated \cite{huang2016linear}. Efficiently extending existing methods to a parallel setting is not straightforward: naive implementation of an Upper Confidence Bound algorithm, for instance, may lead to suboptimal balance between exploration and exploitation when contexts are too similar. The goal of this paper is to formulate methods leveraging the multi-agent structure and to apply them to wireless handover optimization.

The paper is structured as follows. Sections~\ref{sec:def-and-not} and \ref{sec:pb-statement} formally define the parallel contextual bandit problem. Section~\ref{sec:related-work} reviews related methods in the bandit literature. Section~\ref{sec:same-param} develops two approaches for parallel contextual bandits. Finally, Section~\ref{sec:exp} shows empirical performances of our methods, first on a toy example and then on a real wireless base station dataset.

\section{Definitions and notations}
\label{sec:def-and-not}
For any integer $n > 0$, we denote by $[n]$ the set $\{1, \ldots, n\}$ and by $|\mathcal{X}|$ the cardinality of any finite set $\mathcal{X}$. For any $t > 0$ and any sequence $(u_1, u_2, \ldots) \in \mathcal{X}^{\mathbb{N}}$, $\mathbf{u}_t$ will denote the sequence up to index $t$, that is $(u_1, \ldots, u_t)$. For any element $x$ of $\mathbb{R}^d$, $\| x \|$ will denote the euclidean norm and $|x|$ the $\ell_1$-norm. For a given positive definite matrix $A \in \mathbb{R}^{d \times d}$, $\| . \|_A$ denotes the norm induced by the scalar product associated to $A$, that is for any $x \in \mathbb{R}^d$, $\Vert x \Vert_A = \sqrt{x^{\top} A x}$.

\section{Problem statement}
\label{sec:pb-statement}

Let $d > 0$ and let $\mathcal{X} \subseteq \bbR^d$. For a given function $f : \bbR^d \rightarrow \bbR$ and a parameter $\theta_{\star} \in \bbR^d$, we define the reward of a contextual bandit as: for any context $x \in \mathcal{X}$,
\begin{equation}
  \label{def:reward}
  r(x) = f(x; \theta_{\star}) + \varepsilon,
\end{equation}
where $\varepsilon$ is a $R$-sub-gaussian random variable independent of $x$, for some $R > 0$. The function $f$ is called the expected reward. The contextual bandits problem consists in the following. At each iteration $0 < t < T$, a set of contexts $\mathcal{X}_t$ is presented; one aims at selecting the context $x_t \in \mathcal{X}_t$ in order to minimize the expected regret. This boils down to finding the sequence $(x_1, \ldots, x_T) \in \mathcal{X}_1 \times \ldots \times \mathcal{X}_T$ maximizing
\begin{equation}
  \label{eq:expected-regret}
  \sum_{t = 1}^T \mathbb{E}[ r(x_t)] = \sum_{t = 1}^T f(x_t; \theta_{\star}).
\end{equation}
In most settings, the context space $\mathcal{X}$ can be further expanded as a product of a state space $\mathcal{S}$ and an action space $\mathcal{A}$. In other words, at a given iteration, one observes a state $s \in \mathcal{S}$ and wants to find the action maximizing the expected reward:
\[
  a^{\star}(s) \in \argmax_{a \in \mathcal{A}} f(s, a; \theta_{\star}).
\]

When the expected reward is parametrized, a natural strategy is to estimate the parameter $\theta_{\star}$ while limiting the regret as much as possible. This type of methods has been extensively studied in the multi-armed bandit---\emph{i.e.,} context-free bandit---setting \cite{lai1985asymptotically, auer2002finite, bubeck2009online}. More recently, the contextual bandits setting has been investigated, whether the function $f$ is linear \cite{chu2011contextual, agrawal2013thompson}, logistic \cite{chapelle2011empirical, chapelle2015simple} or unknown \cite{agarwal2014taming, srinivas2009gaussian, valko2013finite}.

One way of addressing the problem efficiently is to use an Upper Confidence Bound (UCB) framework. It is a straightforward adaptation from UCB in the multi-armed bandit (MAB) case: given a state $s$ and for each action $a$, one uses past observations to build associated confidence region on the expected reward and chooses the action associated to the greatest possible outcome. The method is formally stated in Figure~\ref{alg:ucb}.

\begin{figure}
  \centering
  \begin{algorithmic}
    \Require{Confidence parameter $\alpha$}
    \For{$t = 1, \ldots, T$}
    \State Receive state $s_t$
    \For{$a \in \mathcal{A}$}
    \State Build a confidence region: $C_{\alpha}(s_t, a)$
    \EndFor
    \State Select $a_t \in \argmax_{a \in \mathcal{A}} \{ \sup \;C_{\alpha}(s_t, a) \}$
    \State Observe reward $r_t$
    \EndFor
  \end{algorithmic}
  \caption{UCB for contextual bandit.}
  \label{alg:ucb}
\end{figure}

In this paper, we focus on the problem where $n > 0$ contextual bandits run in parallel. We consider in addition that they share a similar function $f$, although their parameters $(\theta_{\star}^{(i)})_{1 \leq i \leq n}$ are not necessarily identical. The regret~\eqref{eq:expected-regret} is then replaced by the aggregated regret:
\begin{equation}
  \label{eq:agg-expected-regret}
  \sum_{t = 1}^T \sum_{i = 1}^n f(x_t^{(i)}; \theta_{\star}^{(i)}).
\end{equation}
Without any additional assumption on the parameters $(\theta_{\star}^{(i)})_{1 \leq i \leq n}$, a straightforward strategy is to run one of the aforementioned policies independently on each bandit. However, if the parameters are selected from a restricted set, say $\theta_{\star}^{(i)} \in \Theta$ with $|\Theta| \ll n$, then one may wonder whether it is possible to use this structure to improve the regret minimization policy. This interrogation is investigated in \cite{maillard2014latent} for the multi-armed bandit setting and in \cite{gopalan2016low} for the contextual bandit setting, when the arms (resp. the contexts) are pulled (resp. chosen) sequentially, one at a time. In our setting however, we are interested in finding a strategy for choosing $n$ contexts at each iteration, since the bandits run in parallel. In order to emphasize the interest of our approaches when this assumption is verified, we consider in the remainder of this paper a simpler setting, where every bandit shares the same parameter $\theta_{\star}$ and, consequently, the same expected reward function.


\section{Related work}
\label{sec:related-work}

Surprisingly, the parallel bandit setting has not been widely studied in the contextual bandit case. There is a lot of literature about closely related settings however; we detail each of these settings in the following sections. 

\subsection{Delayed reward}
\label{sec:delayed-reward}

One way to model the parallel bandit problem is to consider bandits with delayed rewards \cite{joulani2013online}: we assume that the reward is not observed immediately after every action but rather delayed. If the rewards are received every $n$ iterations, this setting is then equivalent to $n$ contextual bandits with identical expected reward functions running in parallel.
In the general online learning setting, the delayed feedback is modeled as follows. At a given iteration $t$, the environment chooses a state $s_t$ and a set of admissible actions $\mathcal{A}_t$, just as in the standard setting. The reward however is only observed after a delay $\tau_t$, possibly random, that is usually unknown in advance to the learner. One interesting result, enlightened in \cite[Table 1]{joulani2013online}, is the fact that the additional regret induced by the delay is additive in the stochastic feedback case and multiplicative in the adversarial setting. In our setting, the delay $\tau$ is in the set $\{0, \ldots, n - 1\}$, where $n$ is the number of bandits. Therefore, the additional regret is proportional to the number of bandits.

The general online learning with delayed feedback problem was deeply investigated and can be extended to a wide range of applications (MDPs, distributed optimization, \emph{etc.}), see \eg \cite{joulani2013online} and references therein for further details.

The more specific problem of multi-armed bandits with delayed feedback has been extensively studied in the past decade \cite{guha2010multiarmed, vernade2017stochastic, perchet2016batched}. The particular structure of the problem allows for different approaches than general online learning that sometimes lead to improved convergence guarantees or decreased storage costs.

\subsection{Piled rewards}
\label{sec:piled-rewards}


Our setting offers more structure than a general contextual bandit with delayed rewards since the rewards are accumulated and then all disclosed simultaneously at a given iteration; to the best of our knowledge, this concept of  ``piled rewards'' is only developed in \cite{huang2016linear} for the linear contextual bandit, that is a contextual bandit with a linear reward function:
\[
  f(x; \theta) = x^{\top} \theta.
\]
The proposed algorithm is based on LinUCB, the essential difference being that the covariance matrix is updated as it cycles through the agents. The algorithm is detailed in Figure~\ref{alg:linucb-parallel}.
\begin{figure}
  \centering
  \begin{algorithmic}
    \Require{Confidence parameter $\alpha$}
    \For{$a \in \mathcal{A}$}
    \State Initialize $\hat{A}_a \gets I$, $\hat{b}_a \gets 0$, $\hat{w}_a \gets 0$\;
    \EndFor
    \For{$t = 1, \ldots, T$}
      \For{$i = 1, \ldots, n$}
      \State Select $a^{(i)}_{\star} \in \argmax_{a \in \mathcal{A}} \hat{w}_a^{\top}  s^{(i)}_t + \alpha \| s^{(i)}_t \|_{\hat{A}_a^{-1}}$\;
      \State Update $A_{a^{(i)}_{\star}} \gets A_{a^{(i)}_{\star}} + s^{(i)}_t s^{(i)\top}_t$\;
      \EndFor
      \State Observe rewards $r^{(1)}_t, \ldots, r^{(n)}_t$\;
      \For{$a \in \mathcal{A}$}
      \State Update $\hat{b}_a \gets \hat{b}_a + \sum_{i = 1}^n \mathds{1}_{\{a = a^{(i)}_{\star}\}} r^{(i)}_t s^{(i)}_t$\;
      \State Update $\hat{w}_a \gets \hat{A}_a^{-1} \hat{b}_a$\;
      \EndFor
    \EndFor
  \end{algorithmic}
  \caption{LinUCB-PR algorithm.}
  \label{alg:linucb-parallel}
\end{figure}
This update trick is shown to shrink the confidence area as the different bandits are looped over, that is $\| s \|_{\hat{A}^{-1}}$ does not increase as $\hat{A}$ is updated, for any state $s$. The LinUCB-PR is shown to have a $O(n^2 T |\mathcal{A}| \log(nT|\mathcal{A}| / \delta))$ regret with probability $1 - \delta$. This is not an improvement over LinUCB applied to the parallel setting but it behaves empirically better for large values of $T$, possibly due to the shrinked confidence region. This lack of improvement may be explained by the overconfidence induced by the intermediate updates. As explained in \cite{desautels2014parallelizing}, one should build an overconfidence measure and moderate exploratory redundancy, before deriving any regret bound.

This approach is quite similar to the ones we will detail in the next sections. We will not limit our attention to the linear setting though---the logistic setting will be of particular interest---and we will develop several methods to tackle this problem in a more general fashion.

\subsection{Gaussian processes}
\label{sec:gp}

\begin{figure}
  \centering
  \includegraphics[width=.40\textwidth]{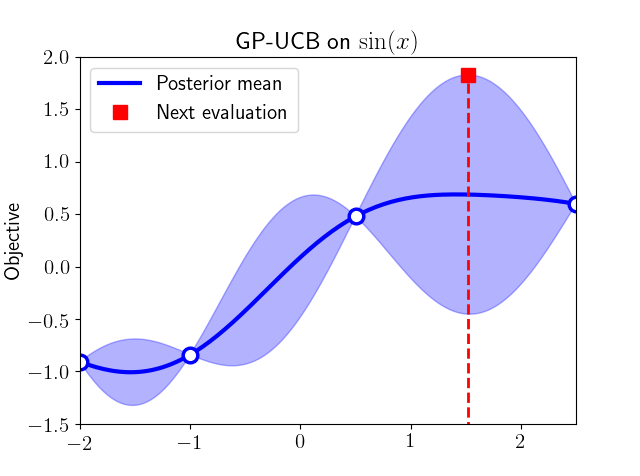}
  \caption{Illustration of GP-UCB algorithm. The posterior mean of the gaussian process is the solid blue line and the associated variance is the filled area. The next evaluation candidate is the one maximizing the \emph{optimistic} prediction of $f$.}
  \label{fig:gp-ucb}
\end{figure}

Lazy optimization with gaussian processes is a particular application of Bayesian optimization where one aims at finding the maximum of a possibly non-convex objective function $f$. The idea is to use a gaussian process as a prior on $f$ and then to sequentially refine the posterior as objective values are observed. A popular method for optimizing in the gaussian process framework is GP-UCB, which is an extension of UCB to gaussian processes. Indeed, at each iteration, one queries the point presenting the highest upper confidence bound based on the posterior of the objective. This is illustrated on Figure~\ref{fig:gp-ucb}.
A typical application lies in hyperparameter tuning for machine learning algorithms \cite{pedregosa2016hyperparameter}. The objective is the negative empirical risk (or any fitting score), which is typically expensive to evaluate. The need to parallelize gaussian process algorithms therefore arose naturally: while being expensive to evaluate, one may have access to additional computational power in order to perform several evaluations simultaneously. However, since GP-UCB selects the optimal candidate in a deterministic fashion, the extension of GP-UCB to a parallel setting is not straightforward. This issue received a lot of attention recently, and many types of approaches have been developed to tackle it. We detail three methods, as they seem to grasp the main ideas of parallelizing, but the reader may find many derivations in, \eg \cite{pmlr-v70-daxberger17a, wang2017batched, gonzalez2016batch, NIPS2015_5804, Haftka2016, NIPS2016_6307}.

First, a method based on pure exploration techniques, namely GP-UCB-PE, has been proposed in \cite{contal2013parallel}. The idea is to select the GP-UCB candidate and define a confidence region around that candidate. Then, the subsequent queries will be selected in order to ``maximize the exploration'', that is each query will select the point in the confidence region with highest posterior variance. Since the posterior variance does not need the objective value to be updated, this ensures distinct candidates among the batch. With a batch size $n$, resulting expected regret is improved by a factor $\sqrt{n}$ in terms of time and is similar for large $T$ in terms of number of queries. This method is somewhat similar to the pure exploration for batched bandits proposed in \cite{perchet2016batched} in the sense that a part of the batch (here $n - 1$ agents) is dedicated to exploring as much as possible in order to guarantee an improved behavior of the remainder of the batch (here the first agent).

Another approach for parallelizing gaussian processes was introduced in \cite{desautels2014parallelizing} as GP-BUCB. As is the case for GP-UCB-PE, this method relies on the fact that posterior variance only depends on the points selected, not their associated values. The next elements are then chosen following a twofold criterion: the standard UCB criterion on the updated process and a overconfidence criterion. The purpose of the latter is to avoid the exploratory redundancy already mentioned in Section~\ref{sec:piled-rewards}---see \cite[Section 4.1]{desautels2014parallelizing} for additional details. Regret analysis of this algorithm yields bounds roughly similar to GP-UCB-PE in terms of number of queries.

Finally, the last method we focus on relies on Thompson sampling---see \eg \cite{hernandez2017parallel, kandasamy2017asynchronous}. At each iteration, the posterior gaussian process is updated. Then, $n$ functions are sampled from this distribution and the $n$ candidates are the maximizers of these sampled functions. This class of methods will be of particular interest in the next section because it seems well-suited for global regret minimization: as opposed to GP-UCB-PE and GP-BUCB, it does not necessarily involve improving the performance of only one or few agents in the batch.

\section{Parallel bandits with identical parameters}
\label{sec:same-param}

In this section, we develop the two main approaches for contextual bandits and see how they can be adapted in the parallel bandit setting.

\subsection{UCB contextual bandits}
\label{sec:ucb-cb}

Upper Confidence Bound (UCB) approaches for contextual bandits rely on the following framework, first introduced for the linear case in \cite{abbasi2011improved}. At a given iteration $t > 0$, one is able to build a confidence region $B_{t - 1}$ for the parameter $\theta_{\star}$, based on the previously selected contexts and associated rewards $(x_1, r_1), \ldots, (x_{t - 1}, r_{t - 1})$. A set of contexts $\mathcal{X}_t$ is proposed and one then chooses the context leading to the best possible reward on $B_{t - 1}$:
\begin{equation}
  \label{eq:oful-rule}
  x_t \in \argmax_{x \in \mathcal{X}_t}\left\{ \max_{\theta \in B_{t - 1}} f(x; \theta) \right\}.
\end{equation}

Methods based on UCB present the advantage of being easy to implement and fast to compute. They obviously require the knowledge of a ``good'' region $B_{t - 1}$ in the sense that $B_{t - 1}$ should be as tight as possible with respect to the selected confidence level.

In the specific case of parallel bandits with identical parameters, the confidence region is based on every bandit historical contexts and rewards: $(x_1^{(i)}, r_1^{(i)}), \ldots, (x_{t - 1}^{(i)}, r_{t - 1}^{(i)})$, for $i \in [n]$. Then, $n$ contexts are selected, one from every $\mathcal{X}_t^{(i)}$, using the UCB rule. Although this method will be preferable to independent policies on each bandits, it may lack the expected exploration/exploitation balance if the contexts sets $\mathcal{X}_t^{(1)}, \ldots, \mathcal{X}_t^{(n)}$ are too similar to enforce different choices amongst the bandits. Indeed, in the---extreme---setting where the contexts sets are identical, the selected contexts will be identical at every bandit: the policy will enforce either a full exploration or a full exploitation scheme, being no different from a setting with only one bandit. The potential regret improvement with respect to independent policies relies solely on the variety of the contexts. This method is formally stated in Figure~\ref{alg:parallel-ucb}

\begin{figure}
  \centering
  \begin{algorithmic}
    \Require{$B_0$, update rule for $B$}
    \For{$t = 1, \ldots, T$}
    \For{$i = 1, \ldots, n$}
    \State Select $x_t^{(i)}$ according to~\eqref{eq:oful-rule}
    \EndFor
    \State Observe the rewards $r_t^{(1)}, \ldots, r_t^{(n)}$
    \State Update $B_t$ with observed rewards and selected contexts
    \EndFor
  \end{algorithmic}
  \caption{UCB algorithm for parallel bandits.}
  \label{alg:parallel-ucb}
\end{figure}

\subsection{Bayesian contextual bandits}
\label{sec:bayesian-cb}

In this section, we focus on bayesian approaches for contextual bandits, and more specifically Thompson-based approaches. In such setting, one defines a prior probability $p(\theta)$ on the parameter to estimate. At iteration $t > 0$, the posterior probability is then of the form
\[
  p(\theta | \mathbf{x}_{t - 1}, \mathbf{r}_{t - 1}) \propto  p(\theta) p(\mathbf{r}_{t - 1} | \mathbf{x}_{t - 1}, \theta),
\]
where $\mathbf{x}_{t - 1} = (x_1, \ldots, x_{t - 1})$, $\mathbf{r}_{t - 1} = (r_1, \ldots, r_{t - 1})$ and $p(\mathbf{r}_{t - 1} | \mathbf{x}_{t - 1}, \theta)$ is the likelihood function. For the sake of simplicity, we use the notation $p(\theta | t - 1)$ for the posterior $p(\theta | \mathbf{x}_{t - 1}, \mathbf{r}_{t - 1})$. Using this relation, one may sample a parameter $\theta_t$ from the posterior probability, either using a closed form or an approximation---\eg MCMC or Laplace approximation. Finally, the context selected from $\mathcal{X}_t$ is the context maximizing the expected reward parametrized by $\theta_t$:
\begin{equation}
  \label{eq:thompson-rule}
  x_t \in \argmax_{x \in \mathcal{X}_t} \left\{ f(x; \theta_t) \right\}.
\end{equation}
Bayesian approaches may offer more flexibility when confidence bounds are not tight but are often much slower to compute, even with rough approximations.

In the specific case of parallel bandits with identical parameters, the posterior is built on every bandit historical contexts and rewards. Then, there are two ways of adapting the regular Thompson approach. First, one may sample one $\theta_t$ and select the contexts $(x_t^{(i)})_{1 \leq i \leq n}$ according to the rule~\eqref{eq:thompson-rule}. Another method is to sample $n$ parameters $\theta_t^{(1)}, \ldots, \theta_t^{(n)}$ independently from the same posterior and then to select every context according to its respective parameter. The former is similar to the adaption of UCB methods to a parallel setting. The latter however benefits from the parallel setting, since it will enforce a better balance between exploitation and exploration at a limited cost (sampling being usually cheap \eg when using Laplace approximation). In the setting where all contexts sets are identical, the randomness of the sampled parameters will preserve the exploitation/exploration balance, even if no reward is observed until every context is chosen. Both approaches are detailed in Figure~\ref{alg:thomp-naive} and \ref{alg:thomp-mult}.

\begin{figure}
  \centering
  \begin{algorithmic}
    \Require{Prior $p(\theta)$}
    \For{$t = 1, \ldots, T$}
      \State Sample a parameter $\theta_t \sim p(\theta | t - 1)$
      \For{$i = 1, \ldots, n$}
        \State Select $x_t^{(i)}$ according to~\eqref{eq:thompson-rule}
      \EndFor
      \State Observe the rewards $r_t^{(1)}, \ldots, r_t^{(n)}$
      \State Update the posterior with observed rewards and selected contexts
    \EndFor
  \end{algorithmic}
  \caption{Naive Thompson-based algorithm for parallel bandits.}
  \label{alg:thomp-naive}
\end{figure}

\begin{figure}
  \centering
  \begin{algorithmic}
    \Require{Prior $p(\theta)$}
    \For{$t = 1, \ldots, T$}
    \For{$i = 1, \ldots, n$}
    \State Sample a parameter $\theta_t^{(i)} \sim p(\theta | t - 1)$
    \State Select $x_t^{(i)}$ according to~\eqref{eq:thompson-rule}
    \EndFor
    \State Observe the rewards $r_t^{(1)}, \ldots, r_t^{(n)}$
    \State Update the posterior with observed rewards and selected contexts
    \EndFor
  \end{algorithmic}
  \caption{Multisampling Thompson-based algorithm for parallel bandits.}
  \label{alg:thomp-mult}
\end{figure}

Previous integrations of Thompson sampling in a parallel setting \cite{hernandez2017parallel, kandasamy2017asynchronous} as well as its well-studied empirical behavior \cite{chapelle2011empirical} suggests that it will behave better---regret-wise---than UCB in the contextual bandit setting. However, the theoretical analysis of such methods is far from trivial and is out of the scope of this paper: even MAB theoretical bounds were provided only a few years ago \cite{agrawal2012analysis, agrawal2013further, kaufmann2012thompson} and bounds for the contextual case are limited to linear payoffs \cite{agrawal2013thompson}. Consequently, the next section is devoted to exhibiting the aforementioned differences between the two methods and then applies both algorithms to the handover parameter tuning on a real base stations dataset.

\section{Experiments}
\label{sec:exp}

We first illustrate on a toy example the advantages, mentioned at the end of the previous section, of the multisampling Thompson-based algorithm over UCB in the case of parallel bandits. We then present the results obtained when applying UCB and Thompson sampling to the online optimization of handover parameters in a wireless cellular network. As explained in the introduction, this problem can be naturally modeled as parallel contextual bandits. In all the experiments, the implemented linear UCB algorithm is the Optimism in the Face of Uncertainty Linear bandit (OFUL) algorithm, described in \cite{abbasi2011improved}.

\subsection{Toy example}
\label{sec:synth}

To illustrate the benefits of the multisampling Thompson-based algorithm described in Figure \ref{alg:thomp-mult} over the UCB algorithm described in Figure \ref{alg:parallel-ucb} we consider the toy example of a linear contextual bandit model. In this case, the expected reward is a linear function of the context $x$ and we assume that the stochastic reward $r$ is given by
\begin{equation}
\label{eq:reward_toy}
r(x, \theta_{\star}) = x^{\top} \theta_{\star} + \varepsilon
\end{equation}
where $\varepsilon \sim \mathcal{N}(0, R^2)$. We also consider here that the context $x = (s, a)$ where $s \in \mathbb{R}^{d-1}$ corresponds to the state received at each iteration $t$ and $a \in \mathbb{R}$ to the action that has to be chosen. This is the case for the wireless cellular network application described in the introduction and section \ref{sec:base_stations}. We take a parameter $\theta_{\star}$ of the form $\theta_{\star} = (\theta^s_{\star}, 1)$, where $\theta^s_{\star} \in \mathbb{R}^{d-1}$ and the true parameter $\theta^s_{\star}$ is sampled from a multivariate Gaussian distribution $\mathcal{N}(0, \mathbf{I}_{d-1})$ and then normalized to a unit norm vector.

At each iteration, a state $s \in \mathbb{R}^{10}$ is sampled from a multivariate Gaussian distribution $\mathcal{N}(0, \sigma_s^2 \cdot \mathbf{I}_{10})$ and the algorithm must choose between one of the 5 actions $a_i=i$, $0 \leq i \leq 4$. The associated reward is generated according to \eqref{eq:reward_toy} where the variance $R^2$ of the noise term $\varepsilon$
is set to $2.5$. For both strategies, multisampling Thompson and UCB, a penalization term $0.01 \cdot \Vert \theta \Vert$ is added to the linear regression.

\subsubsection{Influence of the variance of the states}
\begin{figure}[ht!]
\includegraphics[width=8.5cm]{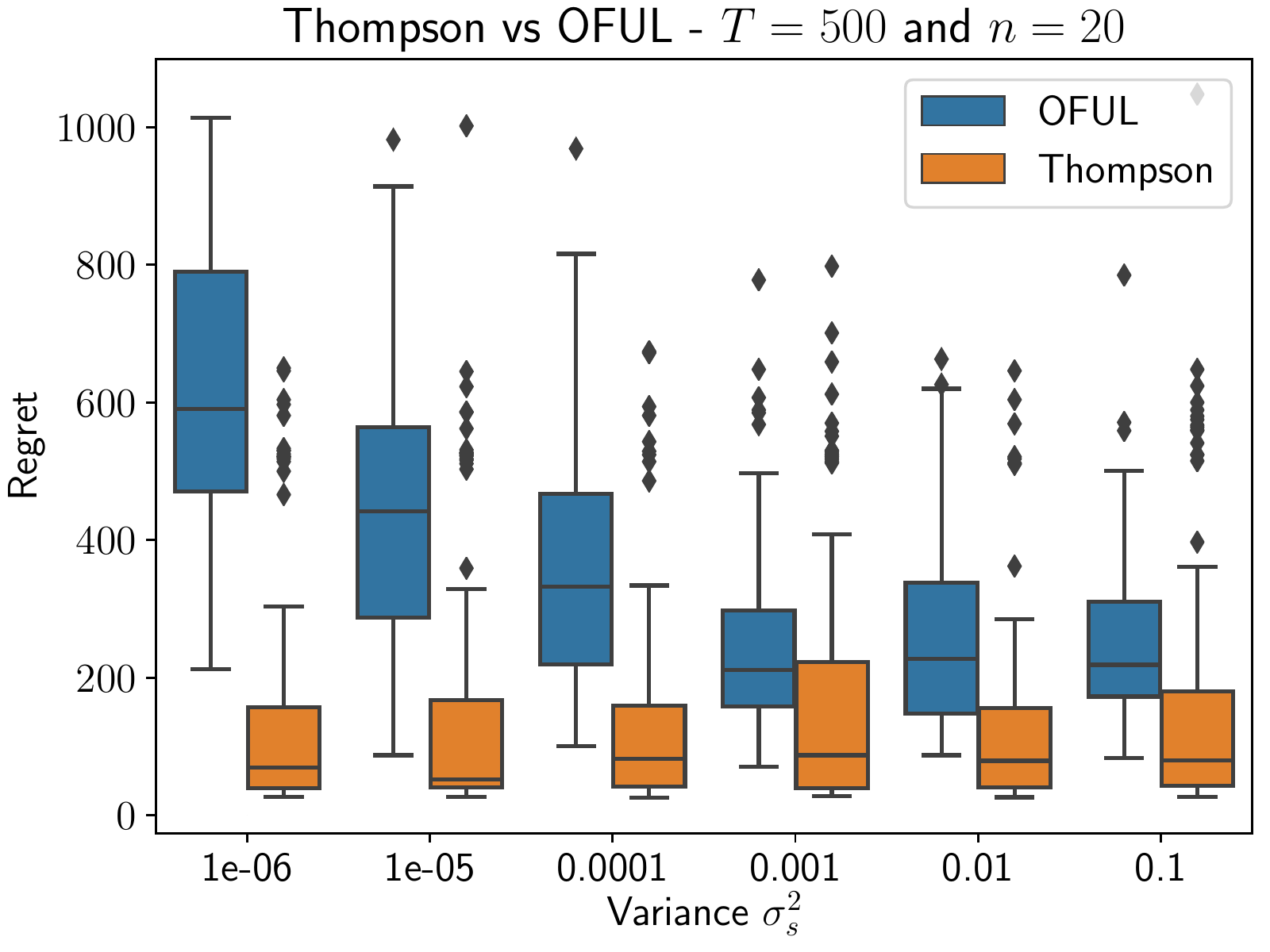}
\caption{Cumulated regrets obtained at a time horizon $T=500$ when running $n=20$ bandits in 
parallel for different values of the variance of the states $\sigma^2_s$.}
\label{fig:thomp_vs_oful_variance}
\end{figure}

We run $n=20$ bandits in parallel and compare the regrets obtained with the multisampling Thompson-based algorithm and the linear UCB algorithm for different variances. The regrets are computed at a time horizon $T=500$ for 100 random repetitions of the algorithm, the randomness coming from the strategies themselves and the generation of the states at each iteration. The results are shown in Figure \ref{fig:thomp_vs_oful_variance}.

One may observe that the regret of UCB decreases as the variance increases. This is explained by the fact that when the variance of the contexts is small, UCB will choose the same action for each of the $n$ bandits that are ran in parallel. This either leads to a full exploration or a full exploitation within each batch of size $n$. On the contrary, the multisampling Thompson-based algorithm is more robust to the change in variance and performs better than UCB. As explained in the previous section, the multisampling Thompson-based algorithm allows for a better balance between exploration and exploitation as $n$ parameters $\theta^{(1)}, \dots, \theta^{(n)}$ are sampled from the posterior $p(\theta | \mathbf{x}_{t - 1}, \mathbf{r}_{t - 1})$.

\subsubsection{Influence of the number of bandits $n$}
\begin{figure}[ht!]
\centering
\includegraphics[width=9cm]{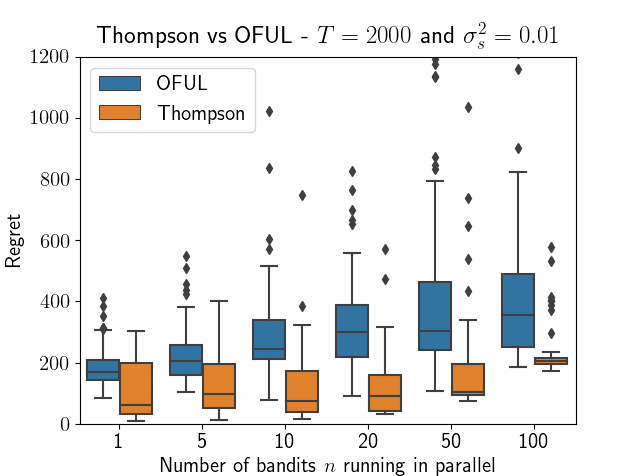}
\caption{Cumulated regrets obtained at time horizon $T=2000$ with a variance of the states $\sigma_s^2 = 0.01$ for different values of the number of bandits ran in parallel $n$.}
\label{fig:thomp_vs_oful_update_every}
\end{figure}

We now study the impact of the number $n$ of bandits ran in parallel. Here, the variance of the states is set to a fixed value $\sigma_s^2 = 0.01$. As above, the regrets are computed at a time horizon $T=2000$ for 100 random repetitions of the algorithm, the randomness coming from the strategies themselves and the generation of the contexts at each step. The results are shown in Figure \ref{fig:thomp_vs_oful_update_every}.

As the number of bandits $n$ increases, the overall regret of UCB degrades whereas the performance of the multisampling Thompson-based algorithm remains stable. This is due to the fact that for large values of the number of bandits $n$, the multisampling Thompson-based algorithm preserves an exploration/exploitation balance compared to UCB.

As explained in section \ref{sec:delayed-reward}, the parallel bandits problem can be seen as bandits with delayed rewards. For the non-contextual case, it was also found in \cite{chapelle2011empirical} that Thompson sampling was more robust to the delay than UCB.

\subsection{Online optimization of base station parameters}
\label{sec:base_stations}

The main motivation of the multisampling Thompson strategy developed in this paper is to tune base station parameters of a cellular wireless network so as to provide a good connectivity for all the users. Here we focus on parameters related to \emph{handovers}. A handover occurs when the connection between a user and a cell is transferred to a neighboring cell in order to ensure the continuity of the radio network coverage and prevent interruptions of communication. The reader can refer to Chapter 2 of \cite{karandikar2017LTEmobility} for an account on handovers. We here only present the different steps of a handover procedure.

\subsubsection{Handovers}

\begin{figure}
\centering
\includegraphics[width=8cm]{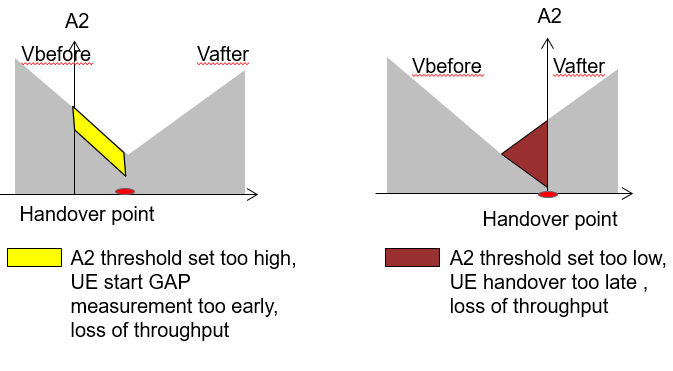}
\caption{Impact of event A2 threshold.}
\label{fig:a2_tradeoff}
\end{figure}

A handover can occur between two cells using the same frequency (\emph{intra-frequency}) or between two cells using different frequencies (\emph{inter-frequency}). For both types of handovers, the user triggers the handover when it receives a better signal from a neighboring cell than from the serving cell. To trigger such an event, the user needs to measure the signal received from neighboring cells. This is automatic for an intra-frequency handover. However inter-frequency measurements are only triggered when the signal received by the serving cell is lower than a pre-specified threshold. Such an event is called \emph{event A2} and is the event of interest in this paper. If the threshold is too low, this results in a late handover and a bad data rate or \emph{throughput} between the cell and the user. On the contrary, if the threshold is too high, unnecessary inter-frequency measurements are triggered and this also results in a bad throughput (Figure \ref{fig:a2_tradeoff}). Indeed, when the user is performing inter-frequency measurements on a neighboring cell, it can no longer exchange information with the serving cell. Tuning the parameter of the base station associated to this threshold can therefore improve the \emph{throughput} between a cell and a user and this should be done for each cell of a wireless network.

Optimization of handover parameters has already been studied in the wireless literature (see e.g.~\cite{isa2015handover,capdevielle2013handover,munoz2013handover,mwanje2014qlearninghandover,sinclair2013handover}). However most of the literature mainly focuses on the optimization of handover performance metrics such as early handovers, late handovers or ping-pong handovers whereas the focus here is on the throughput.

\subsubsection{Data}
\label{sec:data}

Data coming from $n=105$ cells have been recorded every hour during 5 days. For each hour and for each cell the value of the threshold of event A2 and five traffic data features are available. The traffic features are: downlink average number of active users, average number of users, channel quality index of cell edge users, and two features related to the traffic of small data packets. The goal is to recommend the values of the threshold of event A2 so as to achieve the best possible throughput the next hour. As we are only concerned about handovers we aim to maximize the throughput of cell edge users, i.e.~users that are located at the edge of a cell and therefore have a low throughput. We thus define the quantity to optimize as the proportion of users which have a throughput lower than 5 MB/s (Megabytes per second). Obviously, the lower this proportion is the better the parameters are. With the terminology used in this paper and in the contextual bandits literature, the threshold corresponds to the \emph{action}, the quantity that we want to optimize to the \emph{reward} and traffic data features to the \emph{state}.

We emphasize here on the fact that at each hour $t$ we should recommend a threshold value for each one of the 105 cells. Assuming that the reward of each cell is parametrized by a same parameter $\theta$, this is therefore equivalent to running $n=105$ bandits in parallel as described in section \ref{sec:same-param}.

\subsubsection{Parallel logistic contextual bandits}
\label{sec:logistic-bandits}

The reward corresponding to a proportion of users, it appears natural to use a logistic regression model. We therefore recall here the logistic contextual bandit setting. For any $x, \theta \in \bbR^d$, the expected reward is given by $f(x; \theta) = \sigma(x^{\top}\theta)$
where $\sigma : z \mapsto (1 + \exp(-z))^{-1}$ is the sigmoid function and the reward follows a Bernouilli distribution:
\[
  p(r | x, \theta) = f(x; \theta)^{r} (1 - f(x; \theta))^{1 - r},
\]
with $r \in \{0, 1\}$.

Given past observations $(\mathbf{r}_t, \mathbf{x}_t)$,
the penalized negative likelihood estimator is defined by
\[
  \hat{\theta} = \argmin_{\theta} \left\{
    \sum_{s = 1}^t \log(1 + \exp(-r_s \theta^{\top} x_s))
    + \Omega(\theta) \right\}.
\]

From a Bayesian standpoint, if $\Omega = (\lambda_2 / 2) \| \cdot \|^2$ (ridge penalty), the corresponding prior is Gaussian. If $\Omega = \lambda_1 | \cdot | + (\lambda_2 / 2) \| \cdot \|^2$ (elastic net penalty), the corresponding prior is a mixture of a Gaussian distribution and a Laplacian distribution.

Compared to the linear contextual bandits framework, the posterior $p(\theta | t)$ is here intractable. A common way to draw samples from this posterior is to use the Laplace approximation (see \eg section 4.4 in \cite{Bishop2006}): the posterior is approximated by a Gaussian distribution $\mathcal{N}(\mu_t, \Sigma_t)$, with parameters:
\begin{equation}
  \left\{
    \begin{array}{rcl}
      \mu_t &=& \hat{\theta} \\
      \Sigma_t &=& \sum_{s = 1}^t (1 - \sigma(x_s^{\top} \hat{\theta})) \sigma(x_s^{\top}\hat{\theta}) x_s x_s^{\top} \label{eq:cov-mat}.
    \end{array}
  \right.
\end{equation}
In practice, $\Sigma$ can be expensive to compute, so it is common to only use the diagonal coefficients and this is what we do in the experiment.

The assumption that the cells are sharing the same parameter $\theta_{\star}$ may be a bit strong. We show in the next section how to deal with different parameters $\theta^{(i)}_*$, $1 \leq i \leq n$.

\subsubsection{Different parameters}
\label{sec:diff-params}

The parameters $(\theta_{\star}^{(1)}, \ldots, \theta_{\star}^{(n)})$ of each cell are not necessarily identical in practice. However they should benefit from each other's observations. We thus consider that for each $i \in [n]$, $\theta_{\star}^{(i)}$ can be decomposed into the sum of a global parameter $\theta_{\star}$, shared by all the cells, and a local parameter $\tilde{\theta}_{\star}^{(i)}$: $\theta^{(i)}_{\star} = \theta_{\star} + \tilde{\theta}_{\star}^{(i)}$. Let us denote $\boldsymbol{\theta} = (\theta, \tilde{\theta}^{(1)}, \ldots, \tilde{\theta}^{(n)}) \in \bbR^{d \times (n + 1)}$. The new penalized negative likelihood minimizer is:
\begin{equation*}
\begin{aligned}
  \hat{\boldsymbol{\theta}} = \argmin_{\boldsymbol{\theta}} \Biggl\{
    \sum_{s = 1}^t \sum_{i = 1}^n \log(1 + &\exp(-r_s^{(i)} x_s^{(i) \top} (\theta + \tilde{\theta}^{(i)}))) \\
    &+ \Omega(\boldsymbol{\theta}) \Biggr\},
\end{aligned}
\end{equation*}
where again $\Omega$ may be a ridge penalty $\Omega(\boldsymbol{\theta}) = \lambda_2/2 \Vert \theta \Vert^2 + \lambda_2'/2 \sum_{i = 1}^n \Vert \tilde{\theta}^{(i)} \Vert^2$
or an elastic net penalty:
\[
  \Omega(\boldsymbol{\theta}) = \lambda_1 | \theta | + \frac{\lambda_2}{2} \| \theta \|^2 + \lambda_1' \sum_{i = 1}^n | \tilde{\theta} |_1 + \frac{\lambda_2'}{2} \sum_{i = 1}^n \| \tilde{\theta}^{(i)} \|^2.
\]
The diagonal covariance matrix of the Laplace approximation is then:
\[
  \boldsymbol{\Sigma}_t^{diag} =
  \begin{pmatrix}
    \Sigma_t & 0 & \hdots & 0 \\
    0 & \tilde{\Sigma}_t^{(1)} & & \vdots \\
    \vdots & & \ddots & 0 \\
    0 & \hdots & 0 & \tilde{\Sigma}_t^{(n)}
  \end{pmatrix},
\]
where $\Sigma_t$ and the $\tilde{\Sigma}_t^{(i)}$ are defined by substituting the elements of $\hat{\boldsymbol{\theta}}_t$ in~\eqref{eq:cov-mat}.

\subsubsection{Results for wireless handover optimization}

\begin{figure}
\centering
\includegraphics[width=9cm]{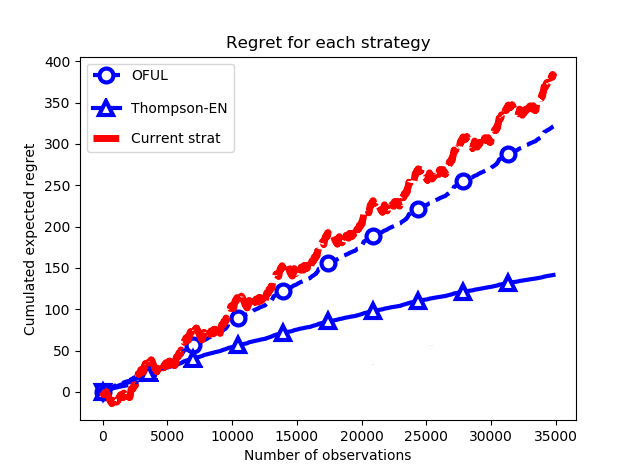}
\caption{Multisampling Thompson-based algorithm versus OFUL for the online optimization of handover parameters. The current strategy denotes the strategy used to collect the data.}
\label{fig:real-exp_regret}
\end{figure}

The Thompsom-based algorithm is applied with the bayesian logistic regression presented above and the OFUL algorithm is applied on the logit transform $\sigma^{-1}$ of the rewards. As the true reward function $r(\cdot, \theta_{\star})$  is unknown, we first fit a logistic regression model on a training data set to learn parameters $(\hat \theta^{(1)}, \dots, \hat \theta^{(n)})$ which we then use as surrogates for the true parameters $(\theta^{(1)}_{\star}, \dots, \theta^{(n)}_{\star})$ in order to evaluate the different strategies: multisampling Thompson-based algorithm, OFUL and the strategy used to collect the data. The cumulated expected regret for each strategy is shown in Figure \ref{fig:real-exp_regret} where one can see that Thompson sampling performs better than OFUL and the strategy used to collect the data.

\section{Conclusion}
\label{sec:conclusion}

In this paper, we exhibited two approaches for handling multi-agent scenarios in the contextual bandits framework. The first one, based on UCB, is a naive extension of the single-agent case; the second one relies on Thompson sampling in order to preserve the exploration-exploitation balance in the bandits batch. Our synthetical experiments enlightened the advantages of Thompson sampling in the parallel setting, as was suggested by theoretical and empirical studies \cite{chapelle2011empirical}. Furthermore, application on wireless handover parameters tuning exhibited a clear superiority of Thompson sampling, in comparison of both manual tuning and UCB-like approach.

Extending this framework to a setting where contextual bandits are not identical but rather regrouped in clusters (as in \cite{maillard2014latent, gopalan2016low}) may be a promising way of generalizing this approach to larger networks. Also, deriving theoretical bounds for Thompson sampling in the parallel setting could lead to additional insights on how to improve existing methods.

\bibliographystyle{ieeetran}
\bibliography{wireless,bandits}






  
\end{document}